\documentclass[aps,pra,showpacs,amsmath,amssymb,amsfonts,tightenlines,twocolumn,lengthcheck]{revtex4-1}
\usepackage{diagbox}
\usepackage{multirow}
\usepackage{amsmath}
\usepackage{amsfonts}
\usepackage{graphicx}
\usepackage{epsfig}
\usepackage{color}
\usepackage{txfonts}
\usepackage[colorlinks,citecolor=blue]{hyperref}

\begin{document}
\title{Generation of two-giant-atom entanglement in waveguide-QED systems}
\author{Xian-Li Yin}
\affiliation{Key Laboratory of Low-Dimensional Quantum Structures and Quantum Control of Ministry of Education, Key Laboratory for Matter Microstructure and Function of Hunan Province, Department of Physics and Synergetic Innovation Center for Quantum Effects and Applications, Hunan Normal University, Changsha 410081, China}
\author{Jie-Qiao Liao}
\email{Corresponding author: jqliao@hunnu.edu.cn}
\affiliation{Key Laboratory of Low-Dimensional Quantum Structures and Quantum Control of Ministry of Education, Key Laboratory for Matter Microstructure and Function of Hunan Province, Department of Physics and Synergetic Innovation Center for Quantum Effects and Applications, Hunan Normal University, Changsha 410081, China}

\begin{abstract}
 We study the generation of quantum entanglement between two giant atoms coupled to a one-dimensional waveguide. Since each giant atom interacts with the waveguide at two separate coupling points, there exist three different coupling configurations in the two-atom waveguide system: separated, braided, and nested couplings. Within the Wigner-Weisskopf framework for single coupling points, the quantum master equations governing the evolution of the two giant atoms are obtained. For each coupling configuration, the entanglement dynamics of the two giant atoms is studied, including the cases of two different atomic initial states: single- and double-excitation states. It is shown that the generated entanglement depends on the coupling configuration, phase shift, and atomic initial state. For the single-excitation initial state, there exists steady-state entanglement for these three couplings due to the appearance of the dark state. For the double-excitation initial state, an entanglement sudden birth is observed via adjusting the phase shift. In particular, the maximal entanglement for the nested coupling is about one order of magnitude larger than those of separate and braided couplings. In addition, the influence of the atomic frequency detuning on the entanglement generation is studied. This work can be utilized for the generation and control of atomic entanglement in quantum networks based on giant-atom waveguide-QED systems, which have wide potential applications in quantum information processing.
\end{abstract}

\date{\today}
\maketitle
\section{Introduction}
Quantum entanglement is a key resource for various quantum information applications~\cite{Einstein35,Horodecki09,Duarte21}, such as quantum key distribution~\cite{Ekert91,Grosshans03}, quantum dense coding~\cite{Bennett92}, quantum teleportation~\cite{Bennett93},  and  quantum-computing technology~\cite{DiVincenzo00}.  The generation of quantum entanglement has been theoretically and experimentally studied in a variety of physical systems, such as optical systems~\cite{Pan12}, trapped ion systems~\cite{Wineland03,Monroe04}, cavity quantum electrodynamics (QED) systems~\cite{Haroche01,Walther06}, circuit-QED systems~\cite{Wallraff21,Blais04,Schoelkopf04}, and waveguide-QED systems~\cite{Roy17,Gu17,Sheremet21}. In particular, the waveguide-QED systems provide an outstanding platform for generating long-distance quantum entanglement and hence they can be regarded as a promising candidate for quantum information processing~\cite{Kimble08,Chang06,Fan07,Law10,Roy11,Fan12,Zoller16,Chang18,Solano20}.

In traditional waveguide-QED systems, the atoms interact with the waveguide at single points and are commonly assumed as pointlike objects, called the dipole approximation~\cite{Wall08}. This approximation is valid in quantum optics when the atoms are assumed to be much smaller than the wavelength of the coupled fields. However, the recent experimental and theoretical advances on giant atoms~\cite{Kockum20Rev} indicate that this approximation becomes invalid when considering the coupling of the superconducting qubits with either the surface acoustic waves (SAWs) or microwave waveguides at multiple coupling points. Typically, quantum interference will take place in coupled quantum systems with multiple coupling points. It has been shown that the quantum interference effect will cause some interesting physical phenomena, such as frequency-dependent Lamb shifts and relaxation rates~\cite{Kockum14PRA}, decoherence-free interatomic interaction~\cite{Kockum18,Oliver20,Ciccarello1,Ciccarello2,Kockum22pra}, unconventional bound states~\cite{Guo20prr,WangX21,Vega21,Wang21,Lim23}, non-Markovian decay dynamics~\cite{Guo17PRA,Delsing19,Longhi20,Du21pra1,Du22A,Yin22A,Lu23}, and single-photon scattering~\cite{Wang20,Du21pra2,Jia21,Liao22}.

Recently, some schemes have been proposed to generate long-range quantum entanglement between distant emitters in traditional waveguide-QED systems~\cite{Vidal11,Baranger13,Porras2013,Ballestero14,Facchi16}. In addition, to increase the maximally achievable entanglement, chiral waveguide setups have also been used to generate two-qubit entanglement~\cite{Zoller14,Moreno15,Schotland16,Mok20}. However, compared to the traditional waveguide-QED systems, quantum interference effects are more abundant and adjustable in giant-atom waveguide-QED systems. Consequently, the generation of entanglement between giant atoms can exhibit new features that will not appear for small atoms~\cite{Wang21pra,Santos23}. Moreover, there exist different coupling configurations between the giant atoms with the waveguide owing to the various arrangements of the coupling points. In this scenario, an interesting question is how the coupling configurations affect the entanglement generation between the two giant atoms.

In this paper we study the generation of quantum entanglement between two giant atoms coupled to a one-dimensional (1D) waveguide with three different coupling configurations: separate, braided, and nested couplings~\cite{Kockum18}. Here, each giant atom couples to the common waveguide at two separate coupling points.  Based on the Wigner-Weisskopf theory~\cite{Scullybook} for single coupling points, we obtain the quantum master equations describing the evolution of the two giant atoms for three different couplings. Note that the quantum master equations in this paper have the same form as that derived through the $(S,L,H)$ formalism in Ref.~\cite{Kockum18}, where $S$ is a scattering matrix, $L$ represents a vector denoting the coupling operators, and $H$ is the Hamiltonian of the system. Concretely, we focus on the entanglement dynamics of the two giant atoms by considering two different initially separable states. For a certain atomic initial state, we find that the entanglement dynamics between the two giant atoms can exhibit different features due to different emission and absorption pathways of photons for these coupling configurations. In the case of the single-excitation initial state,  the maximally achievable entanglement for both the braided and nested couplings can exceed $0.5$. In particular, the entanglement evolution between the two braided giant atoms is characterized by an oscillation in the range from zero to one due to the formation of the decoherence-free interaction. However, for the separate coupling, the maximal entanglement can only reach $0.5$, which is consistent with the small-atom case~\cite{Moreno15}.  In the case of the double-excitation initial state, we can observe the sudden birth of entanglement for these three couplings by adjusting the phase shift. It is shown that the maximally achievable entanglement of the nested coupling is about one order of magnitude larger than those of the other two couplings. We also consider the influence of the atomic frequency detuning on the entanglement generation of the two giant atoms.

The rest of this paper is organized as follows. In Sec.~\ref{Physical model and Eqs} we introduce the physical system for two giant atoms coupled to a common waveguide and present the Hamiltonians. In Sec.~\ref{QMEQ of two GAs} we derive the quantum master equations governing the evolution of the two giant atoms for three different coupling configurations and analyze the mechanism for entanglement generation of the two giant atoms. In Sec.~\ref{EnDys} we study the influence of the quantum interference effect and atomic initial state on the entanglement generation between the two giant atoms. In Sec.~\ref{EnDyschgdelta} we analyze the robustness of our protocol against the atomic frequency detuning. We discuss our results in Sec.~\ref{discussion} and briefly summarize in Sec.~\ref{conclusion}.

\begin{figure}[tbp]
\center\includegraphics[width=0.48\textwidth]{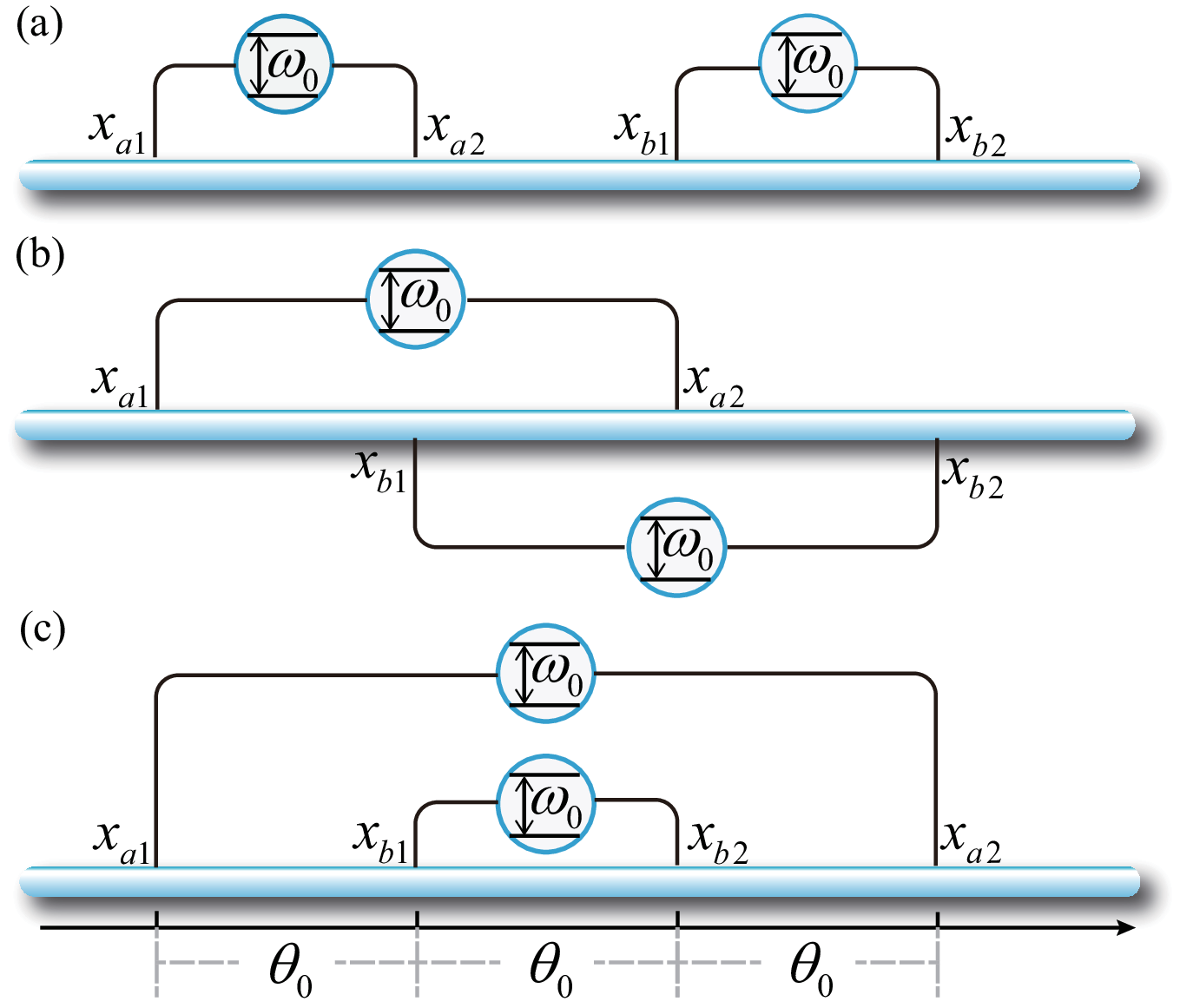}
\caption{Schematic of the coupling configurations for double two-level giant atoms with energy separation $\omega_{0}$ interacting with a common waveguide:  (a) separate,  (b) braided,  and (c) nested couplings. The positions of the coupling points are labeled $x_{jn}$, with $j=a$, $b$ and $n=1$, $2$ referring to the giant atoms and the coupling points, respectively. In all panels, the two giant atoms are initially prepared in two different separable states. The $\theta_{0}=k_{0}d$ is the accumulated phase shift when a single photon with wave vector $k_{0}$ passes through the neighboring coupling points with distance $d$ in the waveguide.}
\label{modelv1}
\end{figure}

\section{System and Hamiltonians}\label{Physical model and Eqs}
We consider a two-giant-atom waveguide-QED system, in which each giant atom interacts with a common 1D waveguide through two separate coupling points, as shown in Fig.~\ref{modelv1}. So far, this system can be realized in experiments on three different platforms: transmon qubits coupled to a SAW waveguide~\cite{Delsing19,Delsing14,Leek17}, Xmon qubits~\cite{Oliver20,Wilson21}, and ferromagnetic spin ensembles~\cite{You22} coupled to a meandering waveguide. By changing the arrangement of the coupling points, there are three different coupling configurations: separated [Fig.~\ref{modelv1}(a)], braided [Fig.~\ref{modelv1}(b)], and nested [Fig.~\ref{modelv1}(c)] couplings. The locations of these coupling points are labeled the coordinates $x_{jn}$, with $j = a$, $b$ marking the giant atoms and $n = 1$, $2$ denoting the two coupling points of each atom. Under the rotating-wave approximation, the Hamiltonian of the system reads ($\hbar=1$)~\cite{Liao22}
\begin{equation}
\hat{H}=\hat{H}_{0}+\hat{H}_{I},
\end{equation}
where
\begin{equation}
\label{freeH}
\hat{H}_{0}=\omega _{0}\sum_{j=a,b}\hat{\sigma}_{j}^{+}\hat{\sigma}_{j}^{-}+\sum_{k}\omega _{k}\hat{c}_{k}^{\dagger }\hat{c}_{k}
\end{equation}
defines the free Hamiltonian of both the two giant atoms and the fields in the waveguide, and
\begin{equation}
\hat{H}_{I}=\sum_{j=a,b}\sum_{n=1,2}\sum_{k}(g_{jn}e^{ikx_{jn}}\hat{c}_{k}\hat{\sigma}_{j}^{+}+\text{H.c.})
\end{equation}
describes the interaction between the giant atoms and the waveguide fields. In Eq.~(\ref{freeH}) $\omega_{0}$ is the transition frequency between the excited state $|e\rangle_{j=a,b}$ and the ground state $|g\rangle_{j}$ of the giant atom $j$. The operator $\hat{\sigma}_{j}^{+}=|e\rangle_{jj}\langle g|$ ($\hat{\sigma}_{j}^{-}=|g\rangle_{jj}\langle e|$) is the raising (lowering) operator of the giant atom $j$ and $\hat{c}_{k}$ ($\hat{c}_{k}^{\dagger}$) is the annihilation (creation) operator of the field in the waveguide with wave vector $k$ and frequency $\omega_{k}$. The constant term $g_{jn}$ is the coupling strength related to the coupling points $x_{jn}$. We use the Wigner-Weisskopf approximation~\cite{Wall08} in the weak-coupling regime and then the coupling strength can be treated as $k$ independent. For simplicity, we consider the case where the coupling strengths at each coupling point are equal to $g$ and the distances between neighboring coupling points are equal to $d$.

In the interaction picture with respect to $H_{0}$, the atom-waveguide coupling Hamiltonian becomes
\begin{equation}
\label{IntHamiltonian}
\hat{V}_{I}(t) =g\sum_{j=a,b}\sum_{n=1,2}\left[ \hat{B}(x_{jn},t) \hat{\sigma}_{j}^{+}e^{i\omega _{0}t}+\hat{B}^{\dagger}(x_{jn},t) \hat{\sigma}_{j}^{-}e^{-i\omega _{0}t}\right],
\end{equation}
where $\hat{B}( x_{jn},t)=[\hat{B}^{\dagger}( x_{jn},t)] ^{\dagger}=\sum_{k}e^{ikx_{jn}}e^{-i\omega_{k}t}\hat{c}_{k}$ is the operator associated with the fields.

\section{Quantum master equations of the two giant atoms}\label{QMEQ of two GAs}
To study quantum entanglement of the two giant atoms, we treat the fields in the waveguide as the environment of the atoms and derive quantum master equations to govern the evolution of the two atoms in three coupling configurations. Under the Markovian approximation, the formal master equation of the system in the interaction picture reads~\cite{Breuer02}
\begin{equation}
\dot{\hat{\rho}}_{I}(t)=-\int_{0}^{\infty }ds\text{Tr}_{w}\left\{ \left[\hat{V}_{I}(t),\left[ \hat{V}_{I}(t-s),\hat{\rho}_{w}\otimes \hat{\rho}_{I}(t)\right] \right] \right\} ,
\end{equation}
where $\hat{\rho}_{I}(t)$ is the density matrix of the two atoms in the interaction picture, $\hat{\rho}_{w}$ is the density matrix of the fields in the waveguide, and $\text{Tr}_{w}\{\bullet\}$ denotes taking the trace over these fields. We consider the case where all the field modes in the waveguide are initially in the vacuum state $\hat{\rho}_{w}=|\emptyset\rangle\langle\emptyset|$, with $|\emptyset\rangle$ representing the empty states. Therefore, we have the relation $\text{Tr}_{w}[ \hat{B}^{\dagger }( x_{in},t) \hat{B}(x_{jn},t-s) \hat{\rho}_{w}] =0$. Using the Wigner-Weisskopf approximation at each single coupling point and assuming $\omega_{k}\approx\omega_{0}+(k-k_{0})\upsilon_{g}$, with $k_{0}$ ($\upsilon_{g}$) the wave vector (group velocity) of the field at frequency $\omega_{0}$~\cite{Fan05,Fan09}, the dynamics of the two giant atoms in these three different coupling configurations are governed by the unified  quantum master equation
\begin{eqnarray}
\label{Meq}
\dot{\hat{\rho}}(t) &=&-i[\hat{H}^{\prime },\hat{\rho}(t)]  \nonumber \\
&&+\sum_{j=a,b}\Gamma _{j}\left\{ \hat{\sigma}_{j}^{-}\hat{\rho}(t)\hat{\sigma}_{j}^{+}-\tfrac{1}{2}[\hat{\sigma}_{j}^{+}\hat{\sigma}_{j}^{-},\hat{\rho}(t)]_{+}\right\}   \nonumber \\
&&+\sum_{i\neq j}\Gamma _{\text{coll}}\left\{ \hat{\sigma}_{i}^{-}\hat{\rho}(t)\hat{\sigma}_{j}^{+}-\tfrac{1}{2}[\hat{\sigma}_{j}^{+}\hat{\sigma}_{i}^{-},\hat{\rho}(t)]_{+}\right\}.
\end{eqnarray}
Hereafter,  we drop the superscript $I$  and always refer to the master equation in the interaction picture. In Eq.~(\ref{Meq}) we introduce the notation $[\bullet,\ast]_{+}\equiv\bullet\ast+\ast\bullet$.
Note that in the derivation of Eq.~(\ref{Meq}) we neglect the propagating time of photons between the coupling points of giant atoms. The Hamiltonian in Eq.~(\ref{Meq}) takes the form
\begin{equation}
\label{DD interaction}
\hat{H}^{\prime }=\sum_{j=a,b}\delta \omega _{j}\hat{\sigma}_{j}^{+}\hat{\sigma}_{j}^{-}+\sum_{i\neq j}g_{ab}\hat{\sigma}_{i}^{+}\hat{\sigma}_{j}^{-},
\end{equation}
where $\delta \omega _{j=a,b}=\sum_{n,m=1,2}\gamma\sin (k_{0}|x_{jn}-x_{jm}|)/2$ is the Lamb shift of the giant atom $j$ and $g_{ab}=\sum_{n,m=1,2}\gamma \sin (k_{0}|x_{an}-x_{bm}|)/2$ is the exchanging interaction strength, with $\gamma =4\pi g^{2}/\upsilon _{g}$ the radiative decay rate of the giant atoms. The parameters $\Gamma _{j=a,b}=\sum_{n,m=1,2}\gamma \cos (k_{0}|x_{jn}-x_{jm}|)$ and $\Gamma _{\text{coll}}=\sum_{n,m=1,2}\gamma \cos (k_{0}|x_{an}-x_{bm}|)$ in Eq.~(\ref{Meq}) are the individual and collective decay rates of the giant atoms, respectively. We note that Eq.~(\ref{Meq}) is consistent with the quantum master equation derived by the $(S,L,H)$ formalism in Ref.~\cite{Kockum18} after returning to the Schr\"{o}dinger picture.

With the quantum master equation~(\ref{Meq}), we can study the entanglement generation for general initial states of the two giant atoms. In this paper we focus on only the cases of initially either one giant atom excited or both giant atoms excited. When the system is restricted to the single-excitation subspace, the jump terms in the quantum master equation~(\ref{Meq}) can be neglected and then a non-Hermitian effective Hamiltonian can be obtained as
\begin{eqnarray}
\label{effH}
\hat{H}_{\text{eff}} &=&\sum_{j=a,b}\delta \omega _{j}\hat{\sigma}_{j}^{+}\hat{\sigma}_{j}^{-}+\sum_{i\neq j}g_{ab}\hat{\sigma}_{i}^{+}\hat{\sigma}_{j}^{-}\nonumber \\
&&-\frac{i}{2}\sum_{j=a,b}\Gamma _{j}\hat{\sigma}_{j}^{+}\hat{\sigma}_{j}^{-}-\frac{i}{2}\sum_{i\neq j}\Gamma _{\text{coll}}\hat{\sigma}_{i}^{+}\hat{\sigma}_{j}^{-}.
\end{eqnarray}
In this case, the evolution of the system can be governed by the Schr\"{o}dinger equation
\begin{equation}
\label{Seq}
i\frac{\partial|\psi (t) \rangle}{\partial t}=\hat{H}_{\text{eff}}| \psi (t) \rangle,
\end{equation}
where $|\psi(t)\rangle$ is the state of the two giant atoms. According to Eq.~(\ref{Seq}), it is straightforward to analytically solve the dynamics of the two giant atoms by assuming their general state in the single-excitation subspace as
\begin{equation}
\label{Tstate}
\left \vert \psi(t)\right \rangle =c_{eg}(t)\left \vert e\right \rangle _{a}\left \vert g\right \rangle _{b}+c_{ge}(t) \left \vert g\right \rangle _{a}\left \vert e\right \rangle _{b},
\end{equation}
where $c_{eg}(t)$ and $c_{ge}(t)$ are the probability amplitudes. Using the Laplace transform and its inverse, we can obtain the analytical expressions of  $c_{eg}(t)$ and $c_{ge}(t)$ under the initial conditions $c_{eg}(0)=1$ and  $c_{ge}(0)=0$.

To study the entanglement generation in the whole state space of the two atoms, we work in the collective state representation, where the two-giant-atom system behaves as a single four-level system with states $|\psi_{2}\rangle=|e\rangle_{a}|e\rangle_{b}$, $|\psi_{0}\rangle=|g\rangle_{a}|g\rangle_{b}$, and $|\psi_{\pm}\rangle$. According to the eigen-equation $\hat{H}^{\prime }|\psi_{\pm }\rangle=E_{\pm }|\psi _{\pm }\rangle$, the expressions of the collective states $|\psi_{\pm}\rangle$ are given by
\begin{equation}
\label{coll states}
|\psi_{\pm }\rangle =N_{\pm }\left( \frac{\delta \omega_{a}-\delta \omega _{b}\pm \Omega }{g_{ab}}|e\rangle_{a}|g \rangle _{b}+2|g\rangle _{a}|e \rangle _{b}\right),
\end{equation}
with the corresponding eigenvalues
\begin{equation}
E_{\pm }=\tfrac{1}{2}(\delta\omega _{a}+\delta\omega _{b}\pm\Omega),
\end{equation}
where $\Omega =\sqrt{4g_{ab}^{2}+(\delta\omega_{a}-\delta\omega_{b})^{2}}$ is the level shift induced by the exchanging interaction and the difference of the Lamb shifts of the two giant atoms. The normalization constants $N_{\pm }$ in Eq.~(\ref{coll states}) are defined by
\begin{equation}
N_{\pm }=\left( 4+\frac{1}{g_{ab}^{2}}(\delta \omega _{a}-\delta \omega_{b}\pm \Omega )^{2}\right) ^{-1/2}.
\end{equation}
\begin{figure}[tbp]
\center\includegraphics[width=0.48\textwidth]{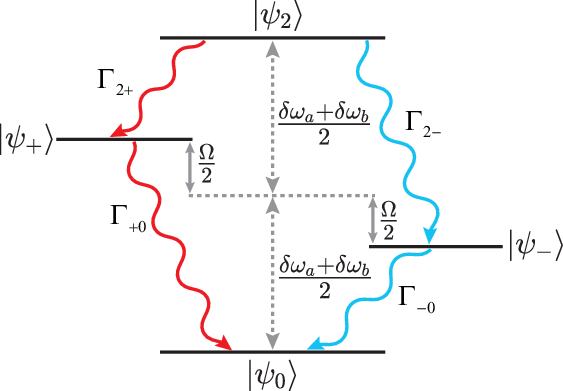}
\caption{Schematic of the levels and decays for the collective states of the two giant atoms. The parameters $\Gamma_{2\pm}$ ($\Gamma_{\pm0}$) describe the transition rates between the collective states $|\psi_{2}\rangle$ ($|\psi_{\pm}\rangle$) and $|\psi_{\pm}\rangle$ ($|\psi_{0}\rangle$). The $\delta\omega_{j=a,b}$ is the Lamb shift of the giant atom $j$ and $\Omega$ is the level shift.}
\label{energylevel}
\end{figure}
Figure~\ref{energylevel} shows the energy-level diagram of the double two-level giant atoms, including the levels and transition rates between different levels. To obtain the transition rates between these levels, we use the basis $\{|\psi_{2}\rangle ,|\psi_{+}\rangle,|\psi_{-}\rangle,|\psi_{0}\rangle \}$ to express the evolution of the diagonal elements of the quantum master equation~(\ref{Meq}) as
\begin{eqnarray}
\label{Eqsforfourlevel}
\dot{\hat{\rho}}_{22}(t)&=&-(\Gamma_{a}+\Gamma_{b})\hat{\rho}_{22}(t),\nonumber\\
\dot{\hat{\rho}}_{++}(t)&=&-i\Delta\lbrack \hat{\rho}_{+-}(t)-\hat{\rho}_{-+}(t)]+\Gamma_{2+}\hat{\rho}_{22}(t)+\Gamma_{++}\hat{\rho}_{++}(t)\nonumber\\
&&+\Gamma_{+-}\hat{\rho}_{+-}(t)+\Gamma_{-+}\hat{\rho}_{-+}(t), \nonumber\\
\dot{\hat{\rho}}_{--}(t)&=&i\Delta\lbrack\hat{\rho}_{+-}(t)-\hat{\rho}_{-+}(t)]+\Gamma_{2-}\hat{\rho}_{22}(t)+\Gamma_{--}\hat{\rho}_{--}(t)\nonumber\\
&&+\Gamma_{+-}\hat{\rho}_{+-}(t)+\Gamma_{-+}\hat{\rho}_{-+}(t),  \nonumber\\
\dot{\hat{\rho}}_{00}(t)&=&\Gamma_{+0}\hat{\rho}_{++}(t)+\Gamma_{+-}\hat{\rho}_{+-}(t)-2\Gamma_{-+}\hat{\rho}_{-+}(t)\nonumber\\
&&-2\Gamma_{-0}\hat{\rho}_{--}(t),
\end{eqnarray}
with
\begin{equation}
\Delta=\frac{(\delta\omega_{a}-\delta\omega_{b})\sqrt{-\eta_{+}\eta_{-}}+2g_{ab}^{2}\left(\sqrt{-\frac{\eta_{-}}{\eta_{+}}}-\sqrt{-\frac{\eta_{+}}{\eta_{-}}}\right)}{2\Omega}.
\end{equation}
According to Eqs.~(\ref{Meq}) and~(\ref{Eqsforfourlevel}), the transition rates can be obtained as
\begin{eqnarray}
\label{D_rates}
\Gamma _{2+} &=&\frac{\Gamma _{b}\alpha _{+}-\Gamma _{a}\alpha _{-}+4g_{ab}\Gamma_{\text{coll}}}{2\Omega },  \nonumber \\
\Gamma _{2-} &=&\frac{\Gamma _{a}\alpha _{+}-\Gamma _{b}\alpha _{-}-4g_{ab}\Gamma_{\text{coll}}}{2\Omega },  \nonumber \\
\Gamma _{+0} &=&-\Gamma _{++}=\frac{\Gamma _{a}\alpha _{+}-\Gamma _{b}\alpha_{-}+4g_{ab}\Gamma _{\text{coll}}}{2\Omega },  \nonumber \\
\Gamma _{-0} &=&-\Gamma _{--}=\frac{\Gamma _{b}\alpha _{+}-\Gamma _{a}\alpha_{-}-4g_{ab}\Gamma _{\text{coll}}}{2\Omega },  \nonumber \\
\Gamma _{+-} &=&\Gamma _{-+}=\frac{\left( \Gamma _{a}-\Gamma _{b}\right)\sqrt{-\alpha _{+}\alpha _{-}}+2g_{ab}\Gamma _{\text{coll}}\left( \sqrt{\frac{-\alpha _{-}}{\alpha _{+}}}-\sqrt{\frac{-\alpha _{+}}{\alpha _{-}}}\right) }{4\Omega }, \nonumber \\
&&
\end{eqnarray}
with $\alpha _{\pm }=( \delta \omega _{a}-\delta \omega _{b}) \pm\Omega$. Equation~(\ref{D_rates}) indicates that the transition rates between the collective states of the two giant atoms depend on the parameters $g_{ab}$, $\delta\omega_{j}$, $\Gamma_{j}$, and $\Gamma_{\text{coll}}$, which can be adjusted by tuning the phase shift $\theta_{0}=k_{0}d$ or designing different coupling configurations.

\begin{figure}[tbp]
\center\includegraphics[width=0.48\textwidth]{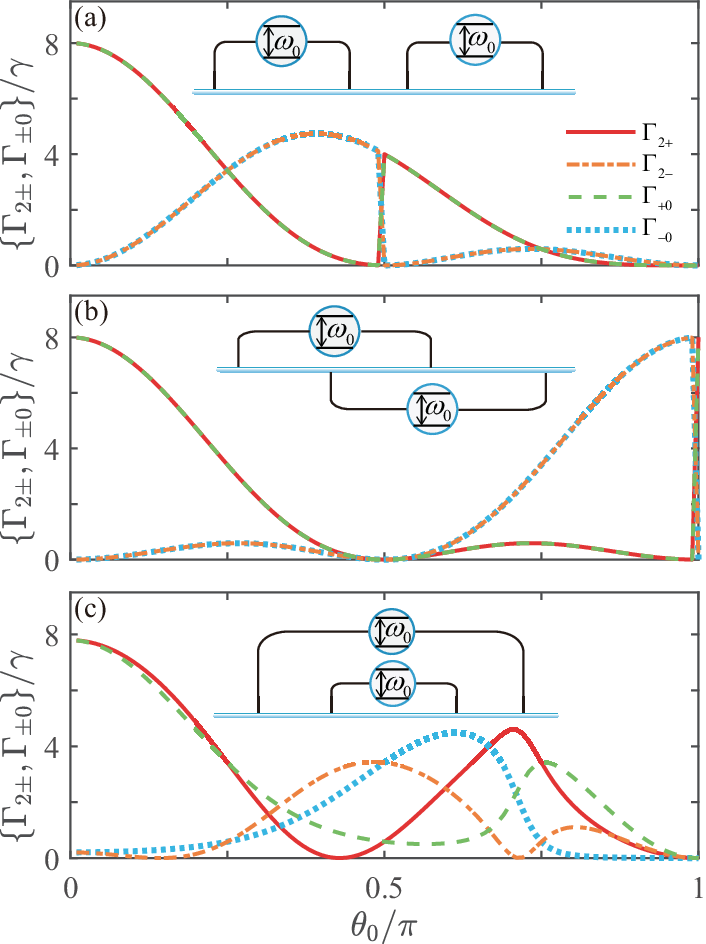}
\caption{Scaled transition rates $\Gamma_{2\pm}/\gamma$ and $\Gamma_{\pm 0}/\gamma$ as functions of the scaled phase shift $\theta_{0}/\pi$ for the (a) separate-,  (b) braided-, and (c) nested-coupling cases.}
\label{transitionrates}
\end{figure}

In Fig.~\ref{transitionrates} we plot the scaled transition rates $\Gamma_{2\pm}/\gamma$ and $\Gamma_{\pm 0}/\gamma$ as functions of the scaled phase shift $\theta_{0}/\pi$ for three coupling configurations. It can be seen that the transition rates satisfy the relations $\Gamma_{2+}=\Gamma_{+0}$ and $\Gamma_{2-}=\Gamma_{-0}$ for both the separate- and braided-coupling cases. However, for the nested-coupling case, the four transition rates have different behaviors. Note that the transition rates for the nested-coupling case have been discussed in Ref.~\cite{Santos23}. Here, to better compare the influence of the coupling configurations on the transition rates, we show the transition rates versus $\theta_{0}$ for these three coupling configurations. Meanwhile, it is straightforward to prove that $\Gamma_{+-}=\Gamma_{-+}=0$ for the separate- and braided-coupling cases, and hence there are no $\hat{\rho}_{+-}(t)$ and $\hat{\rho}_{-+}(t)$ terms in Eq.~(\ref{D_rates}). After obtaining these transition rates, the entanglement generation will be clarified based on Fig.~\ref{energylevel}. In addition, we would like to point out that the system evolves in the absence of the external pumping. To generate the long-lived maximally entangled states, one may drive the double-giant-atom waveguide-QED system with external fields~\cite{Santos23}.


\section{Entanglement dynamics between two giant atoms}\label{EnDys}
In this section, we study the entanglement generation between the two giant atoms for three different coupling configurations shown in Fig.~\ref{modelv1}. To determine the entanglement dynamics of the two giant atoms, we need to solve the quantum master equation~(\ref{Meq}). The entanglement of the two giant atoms can be quantified by the concurrence~\cite{Wootters98}, which is defined as
\begin{equation}
\label{Dfconcurrence}
C(t) =\max (0,\sqrt{\lambda _{1}}-\sqrt{\lambda _{2}}-\sqrt{\lambda _{3}}-\sqrt{\lambda _{4}}),
\end{equation}
where $\lambda_{i}$ are the eigenvalues (in descending order) of the spin-flipped density matrix $\hat{\tilde{\rho}}=\hat{\rho}(\hat{\sigma}_{y}\otimes \hat{\sigma}_{y})\hat{\rho}^{\ast }(\hat{\sigma}_{y}\otimes \hat{\sigma}_{y})$, with $\hat{\sigma}_{y}$ the Pauli spin-flip operator. Note that $C=1$ and $C = 0$ correspond to a maximally entangled state and a separable state, respectively. The quantum master equation~(\ref{Meq}) can be numerically solved under given initial conditions. For each coupling configuration, we will consider two different initial-state cases where the two giant atoms are initially in either the single-excitation state $|\psi(0)\rangle=|e\rangle _{a}|g\rangle _{b}$ or the double-excitation state $|\psi(0)\rangle=|e\rangle _{a}|e\rangle _{b}$.

\subsection{Two-atom entanglement generation in the separate-coupling case}
We begin by considering the separate-coupling configuration, depicted in Fig.~\ref{modelv1}(a). In this case, we can obtain the Lamb shifts $\delta\omega_{a}=\delta\omega_{b}=\gamma\sin\theta_{0}$, the exchanging coupling strength $g_{ab}=\gamma[\sin\theta_{0}+2\sin(2\theta_{0})+\sin(3\theta_{0})]/2$, the individual decay rates $\Gamma_{a}=\Gamma_{b}=2\gamma(1+\cos\theta_{0})$, and the collective decay rate $\Gamma_{\text{coll}}=\gamma[\cos\theta_{0}+2\cos(2\theta_{0})+\cos(3\theta_{0})]$. It can be found that these quantities depend on the phase shift $\theta_{0}=k_{0}d$ between the neighboring coupling points of the giant atoms.  By substituting the expressions of $g_{ab}$, $\Gamma_{a}$, $\Gamma_{b}$, and $\Gamma_{\text{coll}}$ into Eq.~(\ref{D_rates}), the concrete expressions of the transition rates can be obtained and then the physical mechanism for the entanglement generation of the two separate giant atoms can be analyzed.

\begin{figure}[tbp]
\center\includegraphics[width=0.48\textwidth]{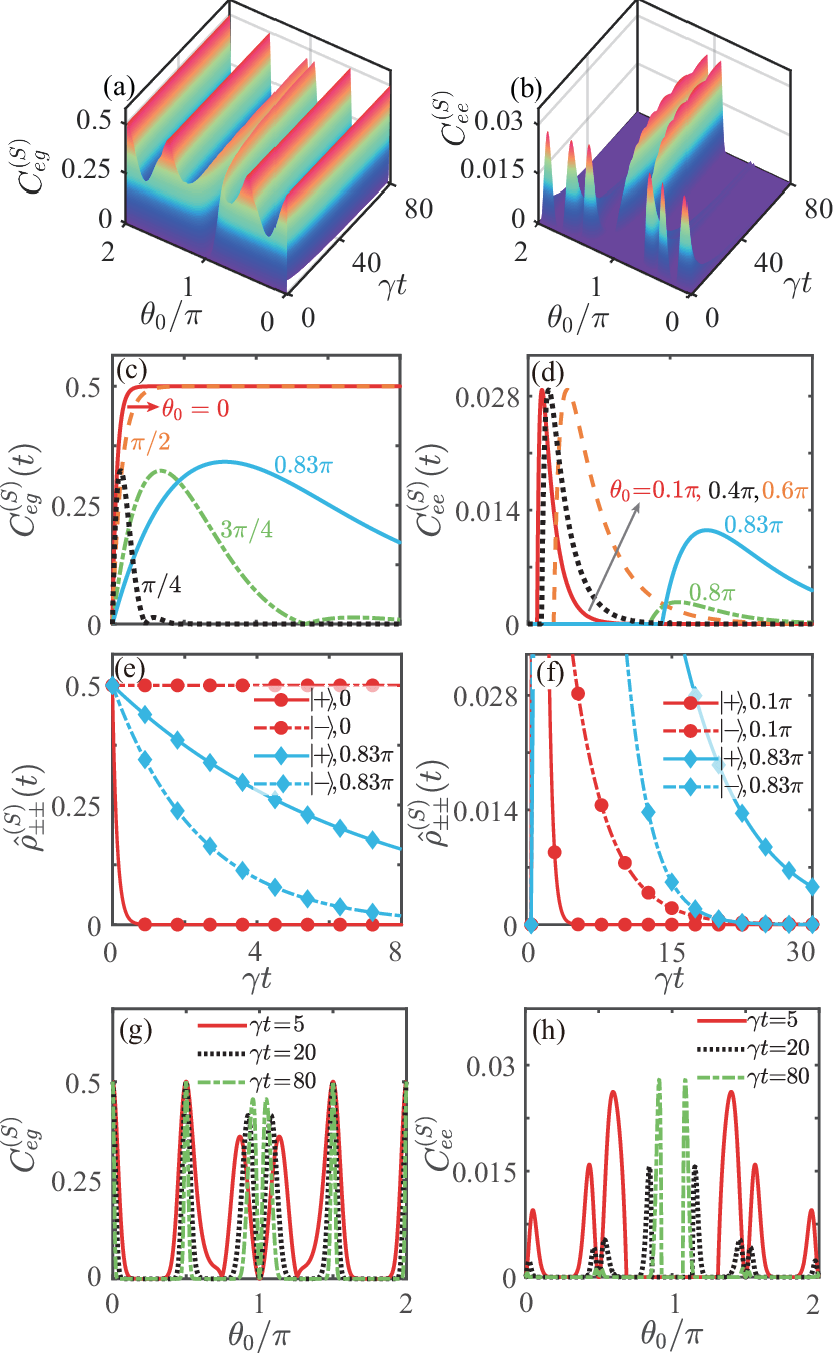}
\caption{Concurrences (a) $C^{(S)}_{eg}$ and (b) $C^{(S)}_{ee}$ in the separate-coupling case as functions of the scaled evolution time $\gamma t$ and the scaled phase shift $\theta_{0}/\pi$, time evolution of (c) $C^{(S)}_{eg}(t)$ and (d) $C^{(S)}_{ee}(t)$ at different values of $\theta_{0}$, (e) and (f) time evolution of the populations $\hat{\rho}_{\pm\pm}^{(S)}(t)$ for the states $|\pm\rangle$, and concurrences (g) $C^{(S)}_{eg}$ and (h) $C^{(S)}_{ee}$ as functions of $\theta_{0}/\pi$ at given values of $\gamma t$. In the left and right columns, the giant atoms are initially in the states $|\psi(0)\rangle=|e\rangle_{a}|g\rangle_{b}$ and $|e\rangle_{a}|e\rangle_{b}$, respectively.}
\label{CS}
\end{figure}

In Figs.~\ref{CS}(a) and~\ref{CS}(b) we show the time evolution of the concurrences $C^{(S)}_{eg}$ and $C^{(S)}_{ee}$, respectively, as functions of the dimensionless quantities $\gamma t$ and $\theta_{0}/\pi$. Note that the superscript $S$ denotes the separate-coupling case and the subscript $eg$ ($ee$) corresponds the atomic initial state $|\psi(0)\rangle=|e\rangle_{a}|g\rangle_{b}$ ($|e\rangle_{a}|e\rangle_{b}$). From Figs.~\ref{CS}(a) and~\ref{CS}(b) we see that both concurrences $C^{(S)}_{eg}$ and $C^{(S)}_{ee}$ are modulated by the phase shift $\theta_{0}$. Meanwhile, the dependence of $C^{(S)}_{eg}$ and $C^{(S)}_{ee}$ on $\theta_{0}$ is a $2\pi$-periodic function. For a phase shift $\theta_{0}\in[0,\pi]$, both $C^{(S)}_{eg}$ and $C^{(S)}_{ee}$  satisfy the relation $C_{eg(ee)}^{(S)}(t,\theta_{0})=C_{eg(ee)}^{(S)}(t,2\pi-\theta_{0})$. It can be found from Fig.~\ref{CS}(a) that there appear six ridges in one period with the increase of time. In particular, when $\theta_{0}=\pi$,  both $C^{(S)}_{eg}$ and $C^{(S)}_{ee}$ are zero. The physical mechanism behind this phenomenon is that the excitation paths of the two giant atoms interfere destructively when $\theta_{0}=\pi$, which results in the exchanging interaction strength $g_{ab}$, individual decay rate $\Gamma_{a}$ ($\Gamma_{b}$), and collective decay rate $\Gamma_{\text{coll}}$ being zero. Thus, the two separate giant atoms are decoupled from the waveguide and there is no entanglement generation between the two giant atoms. To clearly see the effect of the phase shift on the entanglement generation, we further plot in Figs.~\ref{CS}(c) and~\ref{CS}(d) the profiles at typical phase shifts in the region of $\theta_{0}\in[0,\pi]$.

Figure~\ref{CS}(c) depicts the time evolution of the concurrence $C_{eg}^{(S)}(t)$ when $\theta_{0}$ takes typical values. Before analyzing the features in Fig.~\ref{CS}(c), we first present the analytical expression of the concurrence $C_{eg}^{(S)}(t)$. In Sec.~\ref{QMEQ of two GAs} we show that the dynamics of the two giant atoms can be analytically solved when the system is restricted within the single-excitation subspace. For the case of the separate coupling, the concurrence can be analytically calculated as
\begin{equation}
\label{CS_eg}
C^{(S)}_{eg}(t)=e^{-2(1+\cos \theta _{0})\gamma t}\left \vert \sinh \left[4e^{2i\theta _{0}}\cos(\theta_{0}/2)^{2}\gamma t\right] \right \vert.
\end{equation}

By checking the analytical result of Eq.~(\ref{CS_eg}) with the numerical one governed by Eq.~(\ref{Meq}) (not shown here), we find that the dynamics of the two giant atoms can be well described by the effective Hamiltonian when the two giant atoms are initially in the state $|\psi(0)\rangle=|e\rangle_{a}|g\rangle_{b}$. It is straightforward to find that, when $\theta_{0}=0$ and $\pi/2$, the concurrence in Eq.~(\ref{CS_eg}) becomes $C^{(S)}_{eg}(t)=(1-e^{-8\gamma t})/2$ and $C^{(S)}_{eg}(t)=(1-e^{-4\gamma t})/2$, respectively. For the two values of $\theta_{0}$, the concurrence $C^{(S)}_{eg}(t)$ approaches a steady-state value $0.5$ at the rates $8\gamma$ and $4\gamma$, respectively, as shown by the red solid curve and orange dashed curve in  Fig.~\ref{CS}(c). This feature can be explained by the energy-level diagram in Fig.~\ref{energylevel}. For the separate-coupling case, it can be proved that the individual decays and Lamb shifts satisfy the relations $\Gamma_{a}=\Gamma_{b}$ and $\delta\omega_{a}=\delta\omega_{b}$, respectively. In this case, the transition rates $\Gamma_{+-}$ and $\Gamma_{-+}$ in Eq.~(\ref{D_rates}) are zero and the states $|\psi_{\pm}\rangle$ are reduced to the symmetric and antisymmetric states $|\pm\rangle=(|e\rangle_{a}|g\rangle_{b}\pm|g\rangle_{a}|e\rangle_{b})/\sqrt{2}$. When the phase shift is taken as $\theta_{0}=|\epsilon|$ $(\theta_{0}=\pi/2+|\epsilon|)$ with $|\epsilon|\ll1$ in Eq.~(\ref{D_rates}), we have $\Gamma_{+0}\approx 8\gamma$ and $\Gamma_{-0}\approx 0$ $(\Gamma_{+0}\approx 4\gamma$ and $\Gamma_{-0}\approx 0)$. To realize $\Gamma_{+0}\approx 8\gamma$ and $\Gamma_{-0}\approx 0$ $(\Gamma_{+0}\approx 4\gamma$ and $\Gamma_{-0}\approx 0)$, we let $\theta_{0}$ infinitely be close to $0$ and $\pi/2$ but not exactly equal to these two values to ensure $\Omega\neq0$ in Eq.~(\ref{D_rates}). For example, if the phase shift is taken as $\theta_{0}=0.5001\pi$, then we have the transition rate $\Gamma_{-0}=3.95\times10^{-7}\gamma$, which is very small. This allows the population of the state $|\psi_{-}\rangle$ to maintain its initial value 0.5 over a long timescale $1/\Gamma_{-0}\sim10^{6}/\gamma$. Therefore, when $\theta_{0}\rightarrow0$ and $\pi$, the state $|\psi_{-}\rangle$ can be regarded as a dark state decoupled from the waveguide on the concerned timescale. This feature can also be seen from Fig.~\ref{CS}(e), where the population $\hat{\rho}_{--}^{(S)}(t)$ preserves its initial value (see the red dashed curve with circles) while $\hat{\rho}_{++}^{(S)}(t)$ decays to zero (see the red solid curve with circles) with the increase of time. As a result, the concurrence $C^{(S)}_{eg}(t)$ increases gradually with time until it reaches a stationary value $C^{(S)}_{eg}(t\rightarrow\infty)=0.5$.

For the phase shifts $\theta_{0}=\pi/4$ and $3\pi/4$, the concurrence $C^{(S)}_{eg}(t)$ decays to zero fast after achieving the same maximal value [see the black dotted and green dot-dashed curves in Fig.~\ref{CS}(c)]. When $\theta_{0}$ increases to approach $\pi$, we find that $C^{(S)}_{eg}(t)$ exhibits a slower increase over time. After reaching its maximal value, it decays to zero slowly, as shown by the blue solid curve in Fig.~\ref{CS}(c). To explain this phenomenon, we plot the time evolution $\hat{\rho}_{\pm\pm}^{(S)}(t)$ at $\theta_{0}=0.83\pi$ in Fig.~\ref{CS}(e). It can be seen that the population $\hat{\rho}_{--}^{(S)}(t)$ decays to zero fast (see the blue dashed curve with rhombuses), while $\hat{\rho}_{++}^{(S)}(t)$ decreases very slowly (see the blue solid curve with rhombuses). This means that the entanglement generation in this case arises from the population of the state $|\psi_{+}\rangle$.

In the case where the two separate giant atoms are initially in the state $|\psi(0)\rangle=|e\rangle_{a}|e\rangle_{b}$, it can be seen from Fig.~\ref{CS}(b) that the entanglement dynamics exhibits some features different from Fig.~\ref{CS}(a). There are six peaks and two ridges during one period. For this initial state, the two giant atoms first evolve into a mixture of two maximally entangled states $|+\rangle$ and $|-\rangle$ and eventually decay to the ground state $|\psi_{0}\rangle=|g\rangle_{a}|g\rangle_{b}$, as shown by the decay process in Fig.~\ref{energylevel}. In general, the mixture of the states $|+\rangle$ and $|-\rangle$ is not an entangled state. From Fig.~\ref{CS}(a) we see that there is no entanglement generation for all values of $\theta_{0}$ at the initial finite time.  The concurrence $C^{(S)}_{ee}$ is created at later times for some phase shifts, due to the asymmetry between the two transition channels shown in Fig.~\ref{energylevel}. This phenomenon is known as the entanglement sudden (delayed) birth~\cite{Ficek08,Retamal08,Garraway09}.

In Fig.~\ref{CS}(d) we see that the generated entanglement at $\theta_{0}=0.1\pi$, $0.4\pi$, and $0.6\pi$ can reach the same maximal value $C^{(S)}_{ee}(t)\approx0.029$, as shown by the red solid, black dotted, and orange dashed curves, respectively. The magnitude of the maximally achievable entanglement of the two separated giant atoms is only slightly larger than that of the small-atom scheme~\cite{Ficek}. When $\theta_{0}\rightarrow\pi$, the creation of entanglement becomes much later [see the green dot-dashed and blue solid curves in Fig.~\ref{CS}(d)]. In addition, the concurrence $C^{(S)}_{ee}(t)$ is characterized by a slow initial increase followed by a very slow decay. To analyze the contribution of the populations $\hat{\rho}_{\pm\pm}^{(S)}(t)$ on the entanglement generation, we plot the time evolution of $\hat{\rho}_{\pm\pm}^{(S)}(t)$ in Fig.~\ref{CS}(f). It can be seen from Fig.~\ref{CS}(f) that the entanglement generation comes from the population $\hat{\rho}_{--}^{(S)}(t)$ $[\hat{\rho}_{++}^{(S)}(t)]$ when $\theta_{0}=0.1\pi$ $(0.83\pi)$. This means that the entanglement generation is induced by the population of either state $|\psi_{-}\rangle$ or $|\psi_{+}\rangle$ via changing the phase shift.

To clearly see the influence of the quantum interference effect on the entanglement generation of the two giant atoms, we plot the concurrences $C_{eg}^{(S)}$ and $C_{ee}^{(S)}$ as functions of $\theta_{0}/\pi$ at given values of $\gamma t$. In the single-excitation initial state, as shown in Fig.~\ref{CS}(g), the peak values of $C_{eg}^{(S)}$ appear at $\theta_{0}=0$, $\pi/2$, and $\theta_{0}\rightarrow\pi$ within half a period. As the time increases, the peak values of $C_{eg}^{(S)}$ at $\theta_{0}=0$ and $\pi/2$ maintain the steady-state value 0.5 due to the appearance of the dark state. However, the width of the peaks of $C_{eg}^{(S)}$ at $\theta_{0}=\pi/2$ and $\theta_{0}\rightarrow\pi$ becomes narrower. In the double-excitation initial state, there are six peaks in $C_{ee}^{(S)}$, as shown in Fig.~\ref{CS}(h). In addition, the values of the peaks far away from $\theta_{0}=\pi$ decrease gradually while the values of the peaks near $\pi$ gradually increase over time. This feature can be explained from Fig.~\ref{transitionrates}(a), where the transition rates are small and satisfy $\Gamma_{2-}=\Gamma_{-0}>\Gamma_{2+}=\Gamma_{+0}$, resulting in a slower increase and decay of the population of the state $|\psi_{-}\rangle$ compared to that of the state $|\psi_{+}\rangle$. Therefore, the generation of entanglement becomes more delayed, which confirms the results in Fig.~\ref{CS}(d).

\subsection{Two-atom entanglement generation in the braided-coupling case}
We now turn to the case of two braided giant atoms, as shown in Fig.~\ref{modelv1}(b). In this case, we can obtain the Lamb shifts $\delta\omega_{a}=\delta\omega_{b}=\gamma\sin(2\theta_{0})$, the effective exchanging interaction strength $g_{ab}=\gamma[3\sin\theta_{0}+\sin(3\theta_{0})]/2$, the individual decay rates $\Gamma_{a}=\Gamma_{b}=2\gamma[1+\cos(2\theta_{0})]$, and the collective decay rate $\Gamma_{\text{coll}}=\gamma[3\cos\theta_{0}+\cos(3\theta_{0})]$. It has been shown that there exists an interatomic interaction without decoherence at $\theta_{0}=(n+1/2)\pi$~\cite{Kockum18}, with an integer $n$. Below we will show that this kind of interaction enables the entanglement dynamics of the two braided giant atoms to exhibit significant difference from those of the other two coupling configurations.
\begin{figure}[tbp]
\center\includegraphics[width=0.48\textwidth]{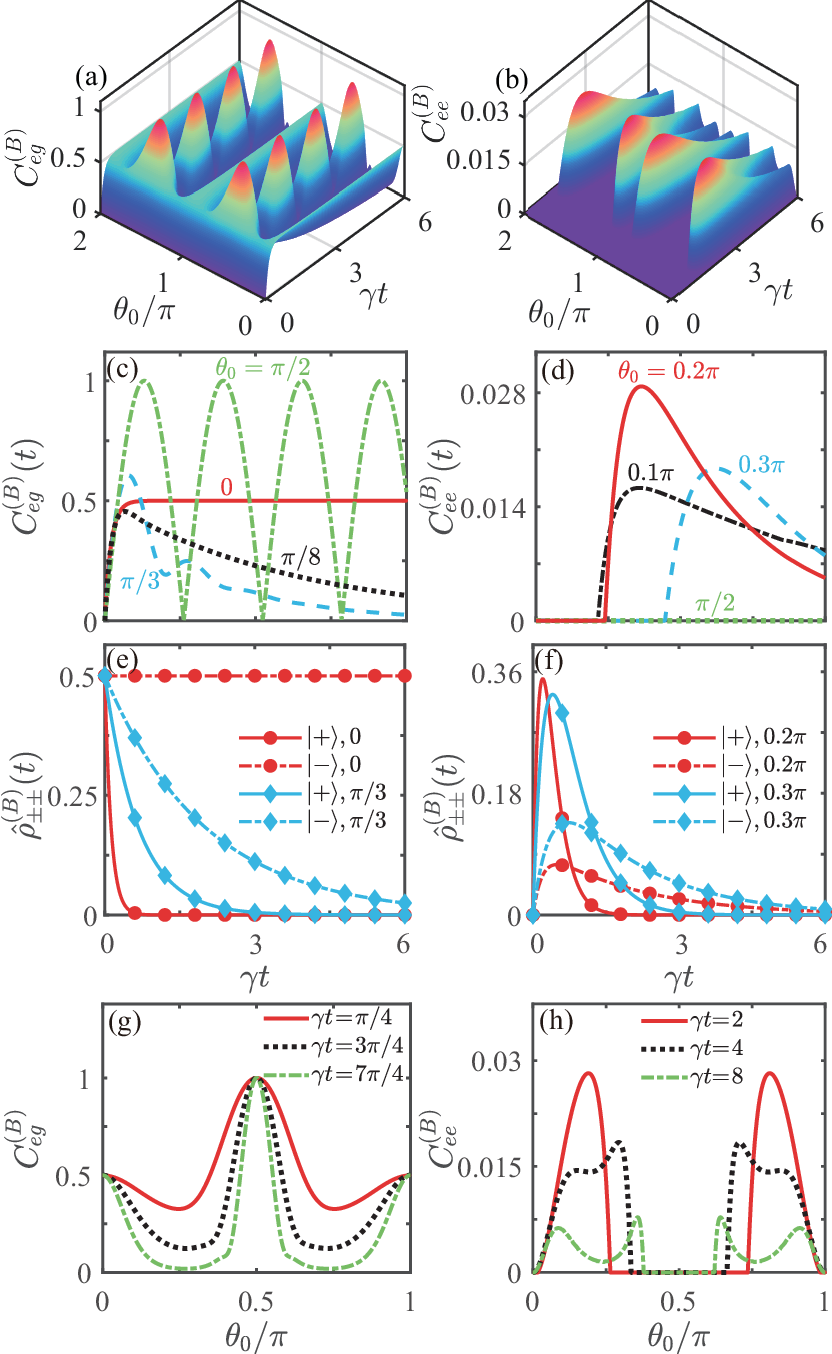}
\caption{Concurrences (a) $C^{(B)}_{eg}$ and (b) $C^{(B)}_{ee}$ in the braided-coupling case as functions of the scaled evolution time $\gamma t$ and the scaled phase shift $\theta_{0}/\pi$, time evolution of  (c) $C^{(B)}_{eg}(t)$ and (d) $C^{(B)}_{ee}(t)$ at different values of $\theta_{0}$, (e) and (f) time evolution of the populations $\hat{\rho}_{\pm\pm}^{(B)}(t)$ for the states $|\pm\rangle$, and concurrences (g) $C^{(B)}_{eg}$ and (h) $C^{(B)}_{ee}$ as functions of $\theta_{0}/\pi$ at given values of $\gamma t$. In the left and right columns, the giant atoms are initially in the states $|\psi(0)\rangle=|e\rangle_{a}|g\rangle_{b}$ and $|e\rangle_{a}|e\rangle_{b}$, respectively.}
\label{CB}
\end{figure}

To study the influence of the quantum interference effect on the entanglement generation, we show the concurrences $C^{(B)}_{eg}$ and $C^{(B)}_{ee}$ versus the dimensionless quantities $\gamma t$ and $\theta_{0}/\pi$ in Figs.~\ref{CB}(a) and~\ref{CB}(b). It can be found that both $C^{(B)}_{eg}$ and $C^{(B)}_{ee}$ are phase dependent with a period of $\pi$, which is different from the separate-coupling case. When the phase shift is in the region of $\theta_{0}\in[0,\pi/2]$, we have the relation $C_{eg(ee)}^{(B)}(t,\theta_{0})=C_{eg(ee)}^{(B)}(t,\pi-\theta_{0})$. Figure~\ref{CB}(a) shows that there are three ridges and $C^{(B)}_{eg}$ exhibits an oscillating process when $\theta_{0}$ is near either $\pi/2$ or $3\pi/2$. Below we show in Figs.~\ref{CB}(c) and~\ref{CB}(d) the profiles at some typical values of the phase shifts in the region $\theta_{0}\in[0,\pi/2]$.

Specifically, the concurrence $C^{(B)}_{eg}(t)$ is characterized by an oscillating process when $\theta_{0}$ is near $\pi/2$ [see the green dot-dashed curve in Fig.~\ref{CB}(c)], whereas it approaches a steady-state value $0.5$ at $\theta_{0}=0$ in the long-time limit [see the red solid curve in Fig.~\ref{CB}(c)].  The oscillation of $C^{(B)}_{eg}$ near $\theta_{0}=\pi/2$ is caused by the nonzero exchanging interaction of the two braided giant atoms. This is because when $\theta_{0}\rightarrow\pi/2$, the collective decay rate and the exchanging interaction strength become $\Gamma_{\text{coll}}\rightarrow0$ and $g_{ab}\rightarrow\gamma$, respectively. In this case, the concurrence $C^{(B)}_{eg}(t)$ is mainly characterized by an oscillating process. However, when $\theta_{0}\rightarrow0$, we obtain $g_{ab}\rightarrow0$ and $\Gamma_{\text{coll}}\rightarrow4\gamma$, which leads to a nonoscillatory contribution to $C^{(B)}_{eg}(t)$. According to these analyses, we know that the concurrence $C^{(B)}_{eg}(t)$ depends on the two parameters $g_{ab}$ and $\Gamma_{\text{coll}}$ in different ways. The oscillation of $C^{(B)}_{eg}(t)$ is caused by the exchanging coupling $g_{ab}$, whereas the nonoscillatory contribution of $C^{(B)}_{eg}(t)$ comes from the collective decay $\Gamma_{\text{coll}}$~\cite{Vidal11}. When $\theta_{0}\neq n\pi/2$, with an integer $n$, we see that $C^{(B)}(t)$ exhibits a fast increase followed by a slow decay [for $\theta_{0}\rightarrow(2n+1)\pi$] or an oscillating decay [for $\theta_{0}\rightarrow(n+1/2)\pi$], as shown by the black dotted and blue dashed curves in Fig.~\ref{CB}(c), respectively.

For the braided coupling, we can also obtain the analytical expression of $C^{(B)}_{eg}(t)$ in the single-excitation subspace. In terms of Eqs.~(\ref{Seq}) and~(\ref{Tstate}) we have
\begin{equation}
\label{CB_eg}
C_{eg}^{(B)}(t)=e^{-4\gamma t\cos ^{2}\theta _{0}}|\sinh [(3e^{i\theta_{0}}+e^{3i\theta _{0}})\gamma t]|.
\end{equation}
From Eq.~(\ref{CB_eg}) we find that when $\theta_{0}=\pi/2$ and $0$, the concurrence is reduced to $C_{eg}^{(B)}(t)=\left\vert \sin (2\gamma t)\right\vert$ and $C^{(B)}_{eg}(t)=(1-e^{-8\gamma t})/2$, respectively. Therefore, $C^{(B)}_{eg}(t)$ exhibits a periodic oscillation in the range from zero to one with a period $\pi/2\gamma$ when $\theta_{0}=\pi/2$, while it tends asymptotically to a steady-state value $0.5$ at a rate $8\gamma$ when $\theta_{0}=\pi$. We note that this feature is consistent with the separate-coupling case. For the braided-coupling case, it can be proved that there also exist equal frequency Lamb shifts and individual decay rates, i.e., $\delta\omega_{a}=\delta\omega_{b}=\gamma\sin(2\theta_{0})$ and $\Gamma_{a}=\Gamma_{b}=2\gamma[1+\cos(2\theta_{0})]$. In this case, the entangled states $|\psi_{\pm}\rangle$ given by Eq.~(\ref{coll states}) become the symmetric and antisymmetric states $|\pm\rangle$. According to Eq.~(\ref{D_rates}), the states $|\pm\rangle$ decay to the ground state $|\psi_{0}\rangle=|g\rangle_{a}|g\rangle_{b}$ with different rates. Therefore, we plot the time evolution of $\hat{\rho}^{(B)}_{\pm\pm}(t)$ in Fig.~\ref{CB}(e) at different values of $\theta_{0}$ when $|\psi(0)\rangle=|e\rangle_{a}|g\rangle_{b}$. For $\theta_{0}=0$, the population $\hat{\rho}^{(B)}_{++}(t)$ decays to zero very fast and $\hat{\rho}^{(B)}_{--}(t)$ remains its maximal value $0.5$ [see the red dashed curve with circles in Fig.~\ref{CB}(e)]. As $\theta_{0}$ increases to $\pi/3$, both the populations $\hat{\rho}^{(B)}_{--}(t)$ and $\hat{\rho}^{(B)}_{++}(t)$ exhibit exponential decay but $\hat{\rho}^{(B)}_{--}(t)$ decreases more slowly [see the blue solid and blue dot-dashed curves with rhombuses in Fig.~\ref{CB}(e)]. Therefore, the entanglement generation is induced by the population of the state $|\psi_{-}\rangle$ at $\theta_{0}=0$ and $\pi/3$.

When the two braided giant atoms are initially in the state $|\psi(0)\rangle=|e\rangle_{a}|e\rangle_{b}$, the concurrence $C_{ee}^{(B)}$ in Fig.~\ref{CB}(b) shows that there are two peaks with one period. In addition, it can be observed that there is no entanglement generation at earlier times, but at some finite times, the entanglement suddenly begins to create for some values of $\theta_{0}$. As time increases, we find that there is no entanglement generation for $\theta_{0}=0$ and $\pi/2$. To explain this phenomenon, we substitute $\theta_{0}=|\epsilon|$ $ (\pi/2+|\epsilon|)$ into Eq.~(\ref{energylevel}) and obtain $\Gamma_{2+}=\Gamma_{+0}\approx 8\gamma$ and $\Gamma_{2-}=\Gamma_{-0}\approx 0$ $(\Gamma_{2+}=\Gamma_{2-}=\Gamma_{+0}=\Gamma_{-0}\approx 0)$. This means that when $\theta_{0}=|\epsilon|$, the decay process of the right channel in Fig.~\ref{energylevel} is forbidden. For the decay process of the left channel, even though it is allowed, the population in the state $|\psi_{2}\rangle=|e\rangle_{a}|e\rangle_{b}$ decays to the ground state $|\psi_{0}\rangle$ very fast. Therefore, there is no entanglement generation. When $\theta_{0}=\pi/2+|\epsilon|$, the decay processes of both the left and right channels are forbidden, and hence the population in the state $|\psi_{2}\rangle$ will not decay into the states $|\pm\rangle$. This is different from the case of the atomic initial state $|\psi(0)\rangle=|e\rangle_{a}|g\rangle_{b}$, as shown by the green dot-dashed curve in Fig.~\ref{CB}(c).

In Fig.~\ref{CB}(d) we plot the time evolution of $C_{ee}^{(B)}(t)$ at selected values of $\theta_{0}$. It can be found that the concurrence can achieve the maximal value $C^{(B)}_{ee}(t)\approx0.029$ by adjusting the values of $\theta_{0}$, which is consistent with the separate-coupling case [see Fig.~\ref{CS}(d)]. However, the values of $\theta_{0}$ corresponding to the maximally achievable $C^{(B)}_{ee}(t)$ and $C^{(S)}_{ee}(t)$ are unequal due to their different coupling configurations. To analyze the mechanism for the entanglement generation in the initial state $|\psi(0)\rangle=|e\rangle_{a}|e\rangle_{b}$, we plot the evolution of the populations $\hat{\rho}_{\pm\pm}^{(B)}(t)$ when $\theta_{0}=0.2\pi$ and $0.3\pi$ in Fig.~\ref{CB}(f), which shows that $\hat{\rho}_{++}^{(B)}(t)$ decays to zero faster than $\hat{\rho}_{--}^{(B)}(t)$. In this case, the entanglement generation is mainly determined by the population of the state $|\psi_{-}\rangle$.

Figures~\ref{CB}(g) and~\ref{CB}(h) show the concurrences $C_{eg}^{(B)}$ and $C_{ee}^{(B)}$ as functions of $\theta_{0}/\pi$ at given values of $\gamma t$. Since $C_{eg}^{(B)}$ and $C_{ee}^{(B)}$ are phase dependent with a period $\pi$, we only consider that the phase shift is in the region of $\theta_{0}\in[0,\pi]$. In the case of the initial state $|e\rangle_{a}|g\rangle_{b}$, the peak values of $C_{eg}^{(B)}$ are located at $\theta_{0}=0$, $\pi/2$, and $\pi$. In the regions of  $\theta_{0}\in(0,\pi/2)$ and $\theta_{0}\in(\pi/2, \pi)$, there exist two dips for $C_{eg}^{(B)}$. According to Eq.~(\ref{CB_eg}), here we take the specific values $\gamma t=\pi/4$, $3\pi/4$, and $7\pi/4$, in which $C_{eg}^{(B)}$ reaches its maximal value 1 at $\theta_{0}=\pi/2$. From Fig.~\ref{CB}(g) we also see that the width of the peaks of $C_{eg}^{(B)}$ becomes smaller for larger values of $\gamma t$, which means that the range of the decoherence-free interaction shrinks. Meanwhile, the width of the dips starts to increase and their maximal values approach zero gradually. In the case of the initial state $|e\rangle_{a}|e\rangle_{b}$, there are two peaks in the concurrence $C_{ee}^{(B)}$ at early time (e.g., $\gamma t=2$). For a larger time, more peaks and smaller peak values appear,  as shown in Fig.~\ref{CB}(h).

\subsection{Two-atom entanglement generation in the nested-coupling case}
Finally, we study the entanglement generation for the nested coupling, as shown in Fig.~\ref{modelv1}(c). In this case, the relevant parameters in the quantum master equation~(\ref{Meq}) are given by $\delta\omega_{a}=\gamma\sin(3\theta_{0})$, $\delta\omega_{b}=\gamma\sin\theta_{0}$, $g_{ab}=\gamma[\sin\theta_{0}+\sin(2\theta_{0})]$, $\Gamma_{a}=2\gamma[1+\cos(3\theta_{0})]$, $\Gamma_{b}=2\gamma(1+\cos\theta_{0})$, and $\Gamma_{\text{coll}}=2\gamma[\cos\theta_{0}+\cos(2\theta_{0})]$. Note that the frequency shifts $\delta\omega_{a}$ and $\delta\omega_{b}$ of the two nested giant atoms are unequal except for $\theta_{0}=n\pi$ and $(2n+1)\pi/4$ with an integer $n$. Compared with the other two coupling configurations, we will see that this feature allows the nested coupling to create greater atomic entanglement in the case of the double-excitation initial state.

\begin{figure}[tbp]
\center\includegraphics[width=0.48\textwidth]{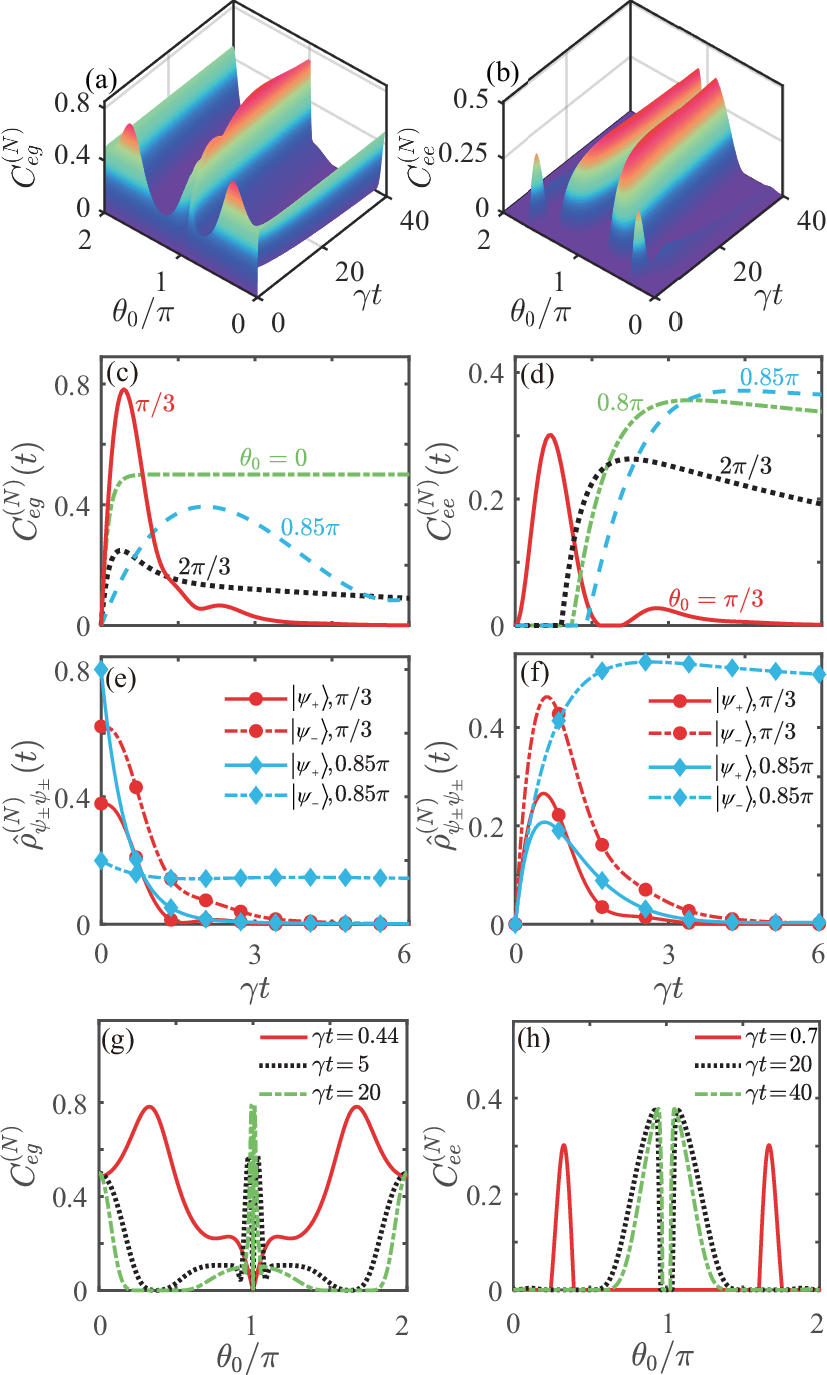}
\caption{Concurrences (a) $C^{(N)}_{eg}$ and (b) $C^{(N)}_{ee}$ in the nested-coupling case as functions of the scaled evolution time $\gamma t$ and the scaled phase shift $\theta_{0}/\pi$,  time evolution of (c) $C^{(N)}_{eg}(t)$ and (d) $C^{(N)}_{ee}(t)$ at different values of $\theta_{0}$, (e) and (f) time evolution of the populations $\hat{\rho}_{\psi_\pm\psi_\pm}^{(N)}(t)$ for the states $|\psi_{\pm}\rangle$, and concurrences (g) $C^{(N)}_{eg}$ and (h) $C^{(N)}_{ee}$ as functions of $\theta_{0}/\pi$ at given values of $\gamma t$. In the left and right columns, the giant atoms are initially in the states $|\psi(0)\rangle=|e\rangle_{a}|g\rangle_{b}$ and $|e\rangle_{a}|e\rangle_{b}$, respectively.}
\label{CN}
\end{figure}

In Figs.~\ref{CN}(a) and~\ref{CN}(b) we show the concurrences $C_{eg}^{(N)}$ and $C_{ee}^{(N)}$, respectively, as functions of the dimensionless quantities $\gamma t$ and $\theta_{0}/\pi$. The atomic initial states are taken as $|\psi(0)\rangle=|e\rangle_{a}|g\rangle_{b}$ and $|e\rangle_{a}|e\rangle_{b}$ in the left and right columns of Fig.~\ref{CN}, respectively. Similar to the separate-coupling case, the concurrences  $C^{(N)}_{eg}$ and $C^{(N)}_{ee}$ are also phase dependent with a period $2\pi$ and satisfy the relation  $C_{eg(ee)}^{(N)}(t,\theta_{0})=C_{eg(ee)}^{(N)}(t,2\pi-\theta_{0})$ for $\theta_{0}\in[0,\pi]$. We find that both $C_{eg}^{(N)}$ and $C_{ee}^{(N)}$ can be created when $\theta_{0}\rightarrow\pi$, as shown by the two ridges in Figs.~\ref{CN}(a) and~\ref{CN}(b). In addition, there are also ridges in Fig.~\ref{CN}(a) when $\theta_{0}=0$ and $2\pi$. Within one period, there exist two peaks in both $C_{eg}^{(N)}$ and $C_{ee}^{(N)}$. To see these features clearly, without loss of generality, we show in Figs.~\ref{CN}(c) and~\ref{CN}(d) the profiles when the phase shift takes typical values in the region of $\theta_{0}\in[0,\pi]$.

Specifically, we see that the concurrence $C^{(N)}_{eg}(t)$ approaches a steady-state value $0.5$ when $\theta_{0}=0$, as shown by the green dot-dashed curve in Fig.~\ref{CN}(c), which is consistent with both the separate- and braided-coupling cases. When the phase shift $\theta_{0}\rightarrow\pi$, the $C^{(N)}_{eg}(t)$ is characterized by a slow initial increase followed by a slow decay, which is the same as the separate-coupling case [see the blue dashed curve in Fig.~\ref{CS}(c)].  However, for the nested coupling, the maximally achievable value of the generated atomic entanglement in Fig.~\ref{CN}(a) exceeds $0.5$ at some values of $\theta_{0}$. For example, it can be seen that the concurrence fast achieves a peak value $C^{(N)}_{eg}\approx0.78$ at $\theta_{0}=\pi/3$, followed by a fast decay to zero. To explain this feature, we take $\theta_{0}=\pi/3$ and then obtain $g_{ab}=\sqrt{3}\gamma$, $\Gamma_{a}=0$, $\Gamma_{b}=3\gamma$, and $\Gamma_{\text{coll}}=0$. In this case, $C^{(N)}_{eg}$ exhibits a fast increase (caused by the nonzero exchanging interaction strength) followed by a fast decay (due to the nonzero individual decay rate of giant atom $b$).

For the nested-coupling configuration, the analytical expression for $C^{(N)}_{eg}(t)$ can be derived from Eqs.~(\ref{Seq}) and~(\ref{Tstate}) as
\begin{equation}
\label{CN_eg}
C_{eg}^{(N)}(t)=\left\vert\frac{F(t,\theta _{0})}{4\cos \left( \frac{\theta _{0}}{2}\right) \sqrt{[(3\cos \theta _{0}-1) ^{2}+4\sin ^{2}\theta_{0}] }}\right\vert,
\end{equation}
where we introduce the function
\begin{eqnarray}
F(t,\theta _{0}) &=&e^{-(A+D)\gamma t}[(1+e^{A\gamma t})(1-e^{B\gamma t})A
\nonumber \\
&&+2ie^{2i\theta _{0}}(1-e^{B\gamma t})(1-e^{B^{\ast }\gamma t})\sin \theta_{0}],
\end{eqnarray}
with
\begin{eqnarray}
\label{N_effecients}
A &=&\sqrt{(5-2e^{i\theta _{0}}+e^{2i\theta _{0}})(e^{i\theta_{0}}+e^{2i\theta _{0}})^{2}},  \notag \\
B &=&\sqrt{8e^{-4i\theta _{0}}\cos ^{2}\left( \frac{\theta _{0}}{2}\right)(3\cos \theta _{0}+2i\sin \theta _{0}-1)},  \notag \\
D &=&2+\cos \theta _{0}+\cos \left( 3\theta _{0}\right) .
\end{eqnarray}
By substituting $\theta_{0}=0$ into Eq.~(\ref{CN_eg}), we have $C_{eg}^{(N)}(t)=(1-e^{-8\gamma t})/2$, which tends to a steady-state value $0.5$ at a rate $8\gamma$ in the long-time limit.

To explain the physical mechanism for the atomic entanglement generation in the nested-coupling configuration, in Fig.~\ref{CN}(e) we plot the time evolution of the populations $\hat{\rho}^{(N)}_{\psi_\pm\psi_\pm}(t)$ of the states $|\psi_{\pm}\rangle$ when the atomic initial state is $|\psi(0)\rangle=|e\rangle_{a}|g\rangle_{b}$. In previous discussion, we have shown that the two giant atoms have equal frequency shifts for the separate and braided couplings.  However, for the nested giant atoms, their frequency shifts are $\delta\omega_{a}=\gamma\sin(3\theta_{0})$ and $\delta\omega_{b}=\gamma\sin\theta_{0}$, respectively, which are unequal except for $\theta_{0}=n\pi$ and $(2n+1)\pi/4$, with an integer $n$. Therefore, the collective states of the two nested giant atoms are described by Eq.~(\ref{coll states}). As shown in Fig.~\ref{CN}(e), the unequal Lamb shifts cause the populations $\hat{\rho}^{(N)}_{\psi_\pm\psi_\pm}(t)$ to evolve from different initial values. When $\theta_{0}=\pi/3$ and $0.85\pi$, the population $\hat{\rho}^{(N)}_{\psi_-\psi_-}(t)$ decays to zero more slowly than $\hat{\rho}^{(N)}_{\psi_+\psi_+}(t)$, and hence the entanglement generation comes from the population of the state $|\psi_{-}\rangle$ when $\theta_{0}$ takes these two phase shifts.

Figure~\ref{CN}(d) shows the profiles of the concurrence $C^{(N)}_{ee}$ when the phase shift $\theta_{0}$ takes some typical values. It can be seen that there is no entanglement generation at earlier times, and entanglement suddenly starts to form at some finite time for $\theta_{0}=2\pi/3$, $0.8\pi$, and $0.85\pi$. When $\theta_{0}=\pi/3$, the concurrence $C^{(N)}_{ee}(t)$ starts to form at $t>0$, which is different from the other two coupling configurations. In particular, we find that the maximally achievable entanglement between the two nested giant atoms initially in the state $|\psi(0)\rangle=|e\rangle_{a}|e\rangle_{b}$ can achieve a maximal value $ C^{(N)}_{ee}(t)\approx0.37$, which is about one order of magnitude larger than those of both the separate and braided couplings. In addition, we find that $C^{(N)}_{ee}(t)$ is characterized by a fast initial increase followed by a slow decay when $\theta_{0}$ approaches $\pi$ at some finite time, as shown by the blue dashed curve in Fig.~\ref{CN}(d). To explain this feature, we substitute $\theta_{0}=\pi-|\epsilon|$ (such as $\theta_{0}=0.85\pi$) into Eq.~(\ref{D_rates}) and obtain $\Gamma_{2+}\approx0.99\gamma$, $\Gamma_{2-}\approx0.91\gamma$, $\Gamma_{+0}\approx1.88\gamma$, and $\Gamma_{-0}\approx0.03\gamma$, which satisfy the relation $\Gamma_{+0}>\Gamma_{2+}>\Gamma_{2-}\gg\Gamma_{-0}$. Therefore, when we adjust the phase shift to approach $\pi$, the decay process from $|\psi_{+}\rangle$ to $|\psi_{0}\rangle$  is faster than that from $|\psi_{2}\rangle$ to $|\psi_{+}\rangle$. However, the decay process from $|\psi_{-}\rangle$ and $|\psi_{0}\rangle$ is much less than that the decay from $|\psi_{2}\rangle$ and $|\psi_{-}\rangle$.

In Fig.~\ref{CN}(d) we plot the time evolution of the populations $\hat{\rho}^{(N)}_{\psi_\pm\psi_\pm}(t)$ of the states $|\psi_{\pm}\rangle$ when the atoms are initially in the double-excitation state $|\psi(0)\rangle=|e\rangle_{a}|e\rangle_{b}$. For the phase shift $\theta_{0}=\pi/3$, the populations $\hat{\rho}^{(N)}_{\psi_\pm\psi_\pm}(t)$ first increase fast with time from zero to their maximal values then decrease fast. For $\theta_{0}=0.85\pi$, the population $\hat{\rho}^{(N)}_{\psi_-\psi_-}(t)$ is characterized by a very slow decay after reaching its maximal value. Therefore, for the two nested giant atoms, the entanglement generation arises from the population in the state $|\psi_{-}\rangle$ when $\theta_{0}\rightarrow\pi$, as shown by the blue dashed curve with rhombuses in Fig.~\ref{CN}(d). This phenomenon is different from the separate-coupling case, where the entanglement generation is mainly determined by the population of the state $|\psi_{+}\rangle$ [see the blue dashed curve with rhombuses in Fig.~\ref{CS}(d)].
\begin{figure}[tbp]
\center\includegraphics[width=0.48\textwidth]{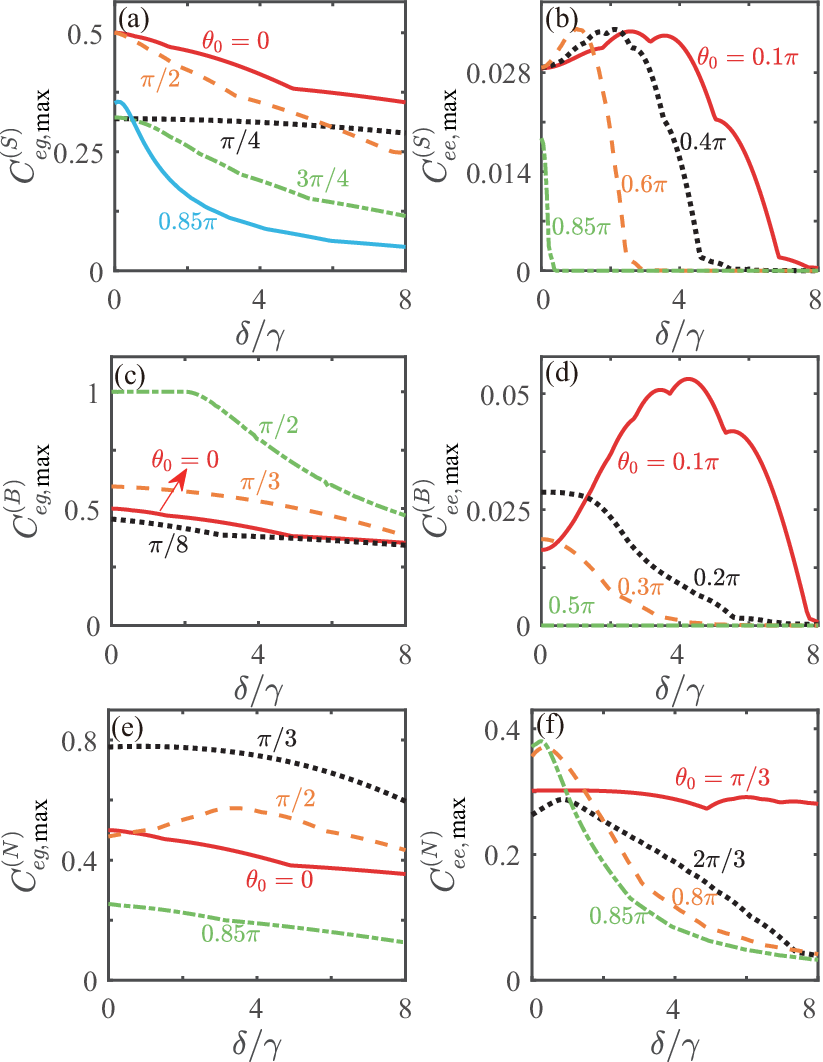}
\caption{Maximal concurrence as a function of the scaled atomic frequency detuning $\delta/\gamma$ for different values of the scaled phase shift $\theta_{0}/\pi$.  In the left and right columns, the giant atoms are initially in the states $|\psi(0)\rangle=|e\rangle_{a}|g\rangle_{b}$ and $|e\rangle_{a}|e\rangle_{b}$, respectively. Two giant atoms in (a) and (b) separate-, (c) and (d) braided-, and (e) and (f) nested-coupling configurations, respectively.}
\label{CSBNMaxvsdelta}
\end{figure}
Figures~\ref{CN}(g) and~\ref{CN}(h) show the concurrences $C_{eg}^{(N)}$ and $C_{ee}^{(N)}$, respectively, as functions of $\theta_{0}/\pi$ at given values of $\gamma t$. In the single-excitation initial state, we find that the concurrence $C_{eg}^{(N)}$ can maintain its peak value 0.5 for larger values of $\gamma t$ when $\theta_{0}=0$ and $2\pi$. In addition, the peak values of $C_{eg}^{(N)}$ can approach 0.8 at both $\theta_{0}=\pi/3$ and $\theta_{0}\rightarrow\pi$. However, the concurrence $C_{eg}^{(N)}$ reaches its peak value more slowly when $\theta_{0}\rightarrow\pi$, as shown by the black dotted and green dot-dashed curves in Fig.~\ref{CN}(g). In the double-excitation initial state, there exist two peaks in $C_{ee}^{(N)}$ within one period when $\gamma t=0.7$. At this value of $\gamma t$, the concurrence  $C_{ee}^{(N)}$ reaches its maximal value when $\theta_{0}=\pi/3$. For larger values of $\gamma t$ (e.g., $\gamma t=20$ and $40$), the concurrence $C_{ee}^{(N)}$ is mainly created at $\theta_{0}\rightarrow\pi$. The physical mechanism behind this phenomenon is consistent with that in Fig.~\ref{CN}(d).

\section{Effect of atomic frequency detuning on the entanglement generation of the two giant atoms}\label{EnDyschgdelta}
In the above discussion, we considered the two resonant-atom case, in which the two atoms have the same energy separation. Since the atomic frequency detuning plays an important role in the induced dipole interaction between the two atoms, in this section we study the influence of the atomic frequency detuning on the entanglement generation of the two giant atoms. To this end, we assume that the resonance frequencies are modified according to $\delta\omega_{a}^{\prime}=\delta\omega_{a}+\delta/2$ and $\delta\omega_{b}^{\prime}=\delta\omega_{b}-\delta/2$ for giant atoms $a$ and $b$, respectively. Figure~\ref{CSBNMaxvsdelta} shows the influence of the frequency detuning $\delta$ between the two giant atoms on the maximal concurrence for different coupling configurations in the cases of two different atomic initial states. Figures~\ref{CSBNMaxvsdelta}(a) and~\ref{CSBNMaxvsdelta}(b), Figs.~\ref{CSBNMaxvsdelta}(c) and~\ref{CSBNMaxvsdelta}(d), and Figs.~\ref{CSBNMaxvsdelta}(e) and~\ref{CSBNMaxvsdelta}(f) correspond to the separate-, braided-, and nested-coupling configurations, respectively. The left and right columns represent that the two giant atoms are initially in the states $|\psi(0)\rangle=|e\rangle_{a}|g\rangle_{b}$ and $|e\rangle_{a}|e\rangle_{b}$, respectively. Numerical results indicate that the maximal concurrence depends on the phase shift, the atomic initial state, the coupling configurations, and the atomic frequency detuning.

For the separate-coupling case, the maximal concurrence $C_{eg,\text{max}}^{(S)}$ in Fig.~\ref{CSBNMaxvsdelta}(a) decreases monotonically as the detuning increases at the selected values of $\theta_{0}$. However, it can be seen that the dependence of $C_{ee,\text{max}}^{(S)}$ at $\theta_{0}=0.1\pi$, $0.4\pi$, and $0.6\pi$ on  the detuning is nonmonotonic [see Fig.~\ref{CSBNMaxvsdelta}(b)]. At these three values of $\theta_{0}$,   $C_{ee,\text{max}}^{(S)}$ first increases with the detuning from the same initial value to their respective maximal values, and then decreases gradually. This is because the decay processes of the left and right channels in Fig.~\ref{energylevel} are modulated by the detuning, which affects the generation of atomic entanglement. When $\theta_{0}=0.85\pi$, as shown by the green dot-dashed curve in Fig.~\ref{CSBNMaxvsdelta}(b), $C_{ee,\text{max}}^{(S)}$ decreases monotonically and becomes fragile with the increase of the detuning.

As shown in Fig.~\ref{CSBNMaxvsdelta}(c), the maximal concurrence $C_{eg,\text{max}}^{(B)}$ exhibits a characteristic of decreasing as $\delta/\gamma$ increases at $\theta_{0}=0$, $\pi/8$, and $\pi/3$. For $\theta_{0}=\pi/2$, we find that $C_{eg,\text{max}}^{(B)}$ retains its maximal value 1 when $\delta/\gamma<2$. In the case where the two braided giant atoms are initially in the double-excitation initial state, it can be seen that there is still no entanglement generation with the increase of $\delta/\gamma$ when $\theta_{0}=\pi/2$.  The concurrence $C_{ee,\text{max}}^{(B)}$ in Fig.~\ref{CSBNMaxvsdelta}(d) decreases monotonically with $\delta/\gamma$ at $\theta_{0}=0.2$ and $0.3\pi$. However,  the $C_{eg, \text{max}}^{(B)}$ is characterized by a nonmonotonic behavior when $\theta_{0}=0.1\pi$.

In the case of the nested coupling, the maximal concurrence $C_{ eg,\text{max}}^{(N)}$ at $\theta_{0}=\pi/2$ is a nonmonotonic function of the detuning, as shown in Fig.~\ref{CSBNMaxvsdelta}(e). This feature is different from the other two coupling configurations. For $C_{ee,\text{max}}^{(N)}$ at $\theta_{0}=\pi/3$ in Fig.~\ref{CSBNMaxvsdelta}(f), it exhibits stronger robustness to the detuning than those at other values of $\theta_{0}$. These results imply that the atomic frequency detuning also plays a significant role in the creation of two-giant-atom entanglement.

\begin{figure}[tbp]
\center\includegraphics[width=0.48\textwidth]{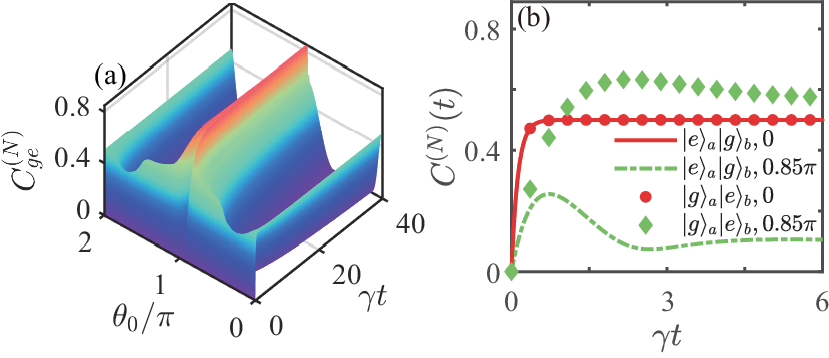}
\caption{(a) Concurrence $C^{(N)}_{ge}$ of the two giant atoms in the nested-coupling case as a function of the scaled evolution time $\gamma t$ and the scaled phase shift $\theta_{0}/\pi$. (b) Time evolution of $C^{(N)}_{eg}(t)$ and $C^{(N)}_{ge}(t)$ at $\theta_{0}=0$ and $0.85\pi$.}
\label{CNgevst}
\end{figure}

\section{Discussion}\label{discussion}

In the single-excitation initial state, we have shown the entanglement generation of two giant atoms in the initial state $|e\rangle_{a}|g\rangle_{b}$. To better understand the influence of the initial state on the entanglement dynamics, below we discuss the entanglement generation when the two giant atoms are initially in the state $|g\rangle_{a}|e\rangle_{b}$. As the two giant atoms satisfy the permutation symmetry in both the separate- and braided-coupling configurations, the entanglement dynamics of the two giant atoms exhibits the same evolution for the initial states $|e\rangle_{a}|g\rangle_{b}$ and $|g\rangle_{a}|e\rangle_{a}$. Therefore, we only need to show the entanglement generation of the two giant atoms initially in $|g\rangle_{a}|e\rangle_{b}$ for the nested coupling. In Fig.~\ref{CNgevst}(a) we plot the concurrence $C^{(N)}_{ge}$ as a function of $\gamma t$ and $\theta_{0}$. By comparing Fig.~\ref{CNgevst}(a) with Fig.~\ref{CN}(a), we find that the concurrences $C^{(N)}_{eg}$ and $C^{(N)}_{ge}$ are characterized by some different features. To clearly see these features, in Fig.~\ref{CNgevst}(b) we show the time evolution of $C^{(N)}_{eg}(t)$ and $C^{(N)}_{ge}(t)$ at $\theta_{0}=0$ and $0.85\pi$. When $\theta_{0}=0$, the concurrences $C^{(N)}_{eg}(t)$ and $C^{(N)}_{ge}(t)$ have the same evolution feature, as shown by the red circles and the red solid curves. However, when $\theta_{0}=0.85\pi$, these two concurrences exhibit different evolution as time increases, as shown by the green rhombuses and the green dot-dashed curves.

Finally, we discuss on the experimental feasibility of our scheme. In Ref.~\cite{Oliver20} it was reported that the two superconducting giant atoms coupled to a common transmission line coplanar waveguide at discrete positions with braided-coupling configuration have been implemented. The coupling strength between the two qubits and the transmission line can reach megahertz in Ref.~\cite{Oliver20}. Hence, the other two coupling configurations are also within the present experimental state of the art. In this work we focus on the Markovian regime by neglecting the photon transmission time. In this regime, the phase shift between the neighboring coupling points of the giant atoms is $\theta_{0}=k_{0}d=2\pi d/\lambda_{\omega}$, where the wavelength $\lambda_{\omega}=2\pi\upsilon_{g}/\omega$ corresponds to the field modes in the waveguide with the giant-atom frequency $\omega=\omega_{0}$. In experiments, since the distance between the coupling points is fixed, the phase shift $\theta_{0}$ can be tuned by changing the giant-atom resonance frequency. When the phase shift is given, we can get the corresponding relation between the typical wavelength of the fields in the waveguide and the distance by using $\theta_{0}=k_{0}d=2\pi d/\lambda_{\omega}$. In our numerical simulations, we take some specifical values of the phase shift. For example, if we take $\theta_{0}=\pi/4$ and $\pi/2$, the corresponding distances between the neighboring coupling points are $\lambda_\omega=8d$ and $4d$, respectively.

In our analysis we neglect both the nonradiative decay and pure dephasing of the giant atoms, because these two decay rates are much smaller than the radiative decay rate in realistic physical systems~\cite{Delsing13,Kirchmair22}. When these two decay rates are taken into account, the entanglement generation will be suppressed. The larger the values of these two decay rates, the smaller the maximally generated entanglement obtained. In the presence of these two decay rates,  the concurrences for these three different coupling configurations will not maintain the steady-state value 0.5 in the long-time limit and eventually the system falls to the ground state. For the braided-coupling configuration, the concurrence $C_{eg}^{(B)}(t)$ at $\theta_{0}=\pi/2$ is characterized by an oscillating decay process due to the suppression of the decoherence-free interaction.
\section{Conclusion}\label{conclusion}
We have studied the entanglement generation between two giant atoms with three different coupling configurations. Using the Wigner-Weisskopf approach for single coupling points, we obtained the quantum master equations governing the dynamics of the two giant atoms. We analyzed the entanglement generation when the two giant atoms are initially in two different separable states. It was shown that the entanglement generation between the two giant atoms depends on the coupling configurations, phase shift, and atomic initial state. Compared to the small-atom scheme, the giant-atom waveguide-QED system can improve the maximally achievable entanglement between two distant emitters. In the case of the single-excitation initial state, the generated entanglement between the two braided giant atoms can reach one when the  decoherence-free interaction is formed by tuning the phase shift. In the case of the double-excitation initial state, the nested coupling can generate the maximal entanglement with magnitude much larger than the other two couplings and the small-atom coupling. We also studied the robustness of our scheme against the changing of the atomic frequency detuning. It was shown that, for different coupling configurations, the maximal concurrence of the two giant atoms exhibits different robustness to the detuning. This work should pave the way for quantum information processing in the giant-atom waveguide-QED systems.

\begin{acknowledgments}
J.-Q.L. was supported in part by National Natural Science Foundation of China (Grants No.~12175061, No.~12247105, and No.~11935006) and the Science and Technology Innovation Program of Hunan Province (Grants No.~2021RC4029 and No.~2020RC4047). X.-L.Y. was supported in part by Hunan Provincial Postgraduate Research and Innovation project (Grant No.~CX20230463).
\end{acknowledgments}

\end{document}